\documentclass[aps, twocolumn, floatfix]{revtex4-1}
\usepackage{amssymb, amsmath, bm, graphicx, color, colortbl, xfrac, tabularx}
\usepackage{todonotes}
\DeclareGraphicsExtensions{.pdf,.png,.jpg, .eps}

\begin{document}
\title{Liquid Time Crystals}

\author{Tapio Simula}
\affiliation{Optical Sciences Centre, Swinburne University of Technology, Melbourne 3122, Australia}

\begin{abstract}
We have constructed and characterised an instrument to study gravitationally bouncing droplets of fluid, subjected to periodic driving force. Our system incorporates a droplet printer that enables an on-demand computer controlled deposition of droplets on the fluid surface. We demonstrate the operation of this instrument by creating and observing long-lived and interacting time crystals whose evolution we have witnessed for more than one hundred thousand oscillation periods. Our observations provide points of comparison for experiments that differentiate between quantum and classical time crystal behaviours in driven non-equilibrium systems.
\end{abstract}

\maketitle

\section{Introduction}
Rain drops falling onto the surface of a lake create a sea of circular surface waves and they may sometimes be seen to skim along the surface before coalescing. The capillary surface waves emerge as part of the gravitational potential energy of the falling droplets is converted into collective degrees of freedom of the fluid. In 1831 Faraday studied an `inverse' scenario by providing the source of energy from below by vibrating a bath filled with a fluid \cite{Faraday1931a}. He found that when the amplitude of the driving increased above a critical value, now called the Faraday threshold, an hydrodynamic instability caused the fluid surface to deform into a non-equilibrium steady state featuring standing Faraday waves. When the fluid is sinusoidally driven with a frequency $f$, the Faraday waves may spontaneously emerge and oscillate at a frequency $f/2$. Floquet analysis of this periodically driven system provides a detailed account on the Faraday wave instability as a subharmonic period doubling bifurcation when the surface tension and inertial effects are brought into competition \cite{Benjamin1954a,Kumar1994a}. Depending on the details of the driving force, a great variety of patterns including triangular, square, and quasicrystalline orderings may be observed \cite{Douady1990a,Edwards1994a}.

In 2005, a team led by Couder studied droplets on the surface of a fluid bath subjected to periodic forcing \cite{Walker1978a} below the Faraday threshold, quantifying the way such droplets began sustained bouncing \cite{Couder2005a,Couder2006a}. Upon changing the driving amplitude, they found the bouncing state to destabilise causing the droplets to begin to `walk' while bouncing on the surface of the fluid. The underlying physical picture is that each impact of the droplet with the fluid surface triggers a capillary wave on the fluid due to the proximity to the Faraday instability. These waves can be long-lived so that the droplet, upon its subsequent bounces, will experience a gradient force exerted by the sloping fluid surface that then drives the planar motion of the droplets. The capillary surface waves periodically added onto the fluid surface by the droplet's impacts coupled to the dynamics of the localised droplet of fluid gives rise to exceptionally rich non-equilibrium dynamics, which has lead to the emergence of a nascent research field coined hydrodynamic quantum analogs. A comprehensive overview of this field is provided in the reviews by Bush \cite{Bush2015a}, and Bush and Oza \cite{Bush2020a}. One curious feature of such driven droplets is that their bouncing period may spontaneously stabilize into a multiple of the driving period via discrete time translation symmetry breaking. 

Goldstein has recently discussed the ubiquity of period multiplying phenomena in nature \cite{Goldstein2018a} including Chladni patterns \cite{Faraday1931a} and swinging pony tails \cite{Keller2010a,Goldstein2012a}. There has been a dramatic increase in the interest in such driven non-equilibrium phenomena since Wilczek coined the concept of a time crystal \cite{Wilczek2012a,Shapere2012a}. Although the original idea of a ground state emerging via a spontaneous time translation symmetry breaking was soon proved to be an impossibility \cite{Bruno2013a,Watanabe2015a}, a related concept where explicitly broken time translation symmetry is reduced to a lower discrete symmetry in a periodically forced non-equilibrium system was soon resurrected by Sacha \cite{Sacha2015a}. Sacha's original concept \cite{Sacha2015a,Sacha2018a} considered a cloud of interacting Bose--Einstein condensate (BEC) bouncing on a periodically driven atom mirror, where the centre of mass of the BEC would begin to oscillate at a subharmonic frequency $f/2$ with respect to that of the mirror driving frequency $f$. An experimental effort of building such time crystals is currently underway \cite{Giergiel2020a}. In our experiments the mirror made of sheet of light vibrated at frequencies around $10,000$ Hz is replaced by a mirror made of sheet metal vibrated at a frequency of around $100$ Hz and the micrometer scale atom cloud is replaced by a millimetric polymer droplet.

The first part of this paper is devoted to description of the key components of our driven droplets instrument. We begin by providing an overview of its main parts in Sec.~\ref{layout} by characterising the mechanical properties of the electrodynamic shaker with (Sec.~\ref{shaker}) and without (Sec.~\ref{bearing}) a stabilising air bearing. Section~\ref{bath} provides the details of the fluid containing bath. We then describe the droplet printer (Sec.~\ref{printer}) that enables an on demand computer controlled deposition of the droplets onto the fluid surface. Most of the information drawn from our experiments is through optical imaging and subsequent image processing, described in Sec.~\ref{imaging}. Section~\ref{layout} ends with a discussion of the control systems (Sec.~\ref{control}) used for operating the experiments. Section \ref{bearingtest} describes the compliance characteristics of the driving system of the instrument in the absence of fluid. Thermal characterisation of the system is discussed in Sec.~\ref{thermal}. In the second part of this paper, we deploy the instrument for demonstrating time crystalline behaviours. Section \ref{results} begins with the creation of Faraday waves (Sec.~\ref{faraday}). Section \ref{bouncing} focuses on creation and observation of bouncing droplets and the last part (Sec.~\ref{manybody}) of the section highlights interesting many-body effects observed in lattices made of superwalking droplets \cite{Valani2019a,Valani2021a}. We close the paper with a discussion in Sec.~\ref{conclusions}.

\section{Details of the instrument}
\label{layout}

The overall design of our experiment is based on the air bearing stabilized shaker system developed by Harris and Bush \cite{Harris2015a,Harris2015b}. The key components of our instrument are constructed on and around a custom optical table (Thorlabs Nexus, 210 mm thickness, 262 kg weight) resting upon passive vibration isolation legs (Thorlabs PTP703). The manufacturer specifies a maximum dynamic deflection coefficient of $0.4\times 10^{-3}$, maximum relative tabletop motion of $0.14$ nm, and deflection (stiffness) of $1.7\;\mu$m under $150$ kg load for this optical table. In addition to providing isolation from external noise sources, the table provides ergonomic benefits due to its $910$ mm working height. The optical table is surrounded by a custom built aluminium extrusion frame that provides a multilevel support structure for appliances and cables, and an easily upgradable enclosure for laser safety purposes. The top shelf is used for storing heat generating electronics such as power supplies and helps in keeping dust away from the fluid bath. 

\subsection{Electrodynamic shaker}\label{shaker}
The driving of our system is achieved using the tried and tested \cite{Harris2015a} electrodynamic shaker (DataPhysics V55/PA300E) weighing $42\;$kg and which is bolted onto a layer cake structure comprising two ($400$ x $200$ x $100$)~mm granite blocks sandwiched between two $32$~mm thick steel plates. The steel plates and the granite blocks were cost-effectively re-purposed from old physics experiments of Bartlett et al. \cite{Bartlett1975a} and Lau et al. \cite{Lau1999a}, respectively. The layer cake platform weighs an estimated $150\;$ kg and is resting on four rubber padded, height adjustable, machine mounts (Sunnex OSM M1). The shaker is aligned with the local gravitational field with the help of a two-axis NIST traceable digital level (DigiPas DWL-1500XY) having an accuracy of $\pm 0.002$ degrees.

\subsection{Air bearing}\label{bearing}
Following Harris and Bush \cite{Harris2015a}, we use similar custom mounts and a $140\;$mm long stainless steel drive rod of radius  $1.0\;$ mm to couple the shaker to the lower end of the square slider bar (length $325$ mm, cross section $50$~mm x $50$~mm, mass 2.2 kg) of the air bearing (OAV BX5050). The air bearing goes through a $152\;$mm diameter custom made port hole in the optical table. The air bearing carriage is fixed rigidly on the underside of a ($400$ x $400$ x $20$)~mm aluminium plate, which is mounted onto the top surface of the optical table using three height adjustable steel posts (Thorlabs BLP01/M). The first resonance frequency of the aluminium plate is well above the experimentally relevant frequencies and does not interfere with our measurements. After the fluid bath, discussed in Sec.~\ref{bath}, has been leveled the height of the posts are locked and precision adjusters are used for translating the air bearing assembly horizontally to bring it in alignment with the shaker. The air bearing assembly is then clamped (Thorlabs PF175B) in place from the feet of the three posts before the air bearing slider bar is lowered and connected to the drive rod with grub screws. We supply the air bearing with a high purity compressed air with $0.01 \mu $m particulate size and maximum oil carryover efficiency of $0.003$ mg/m$^3$ (Walker Filtration). The air line supply pressure is set to $450$ kPa unless stated otherwise. 

\subsection{Fluid bath}\label{bath}
The fluid bath is made of black anodised aluminium and has an outer diameter of $130\;$mm, fluid containing diameter of $100\;$mm with $16\;$mm wall height, and a total mass of $570$~g including mounting screws. It is mounted on top of the air bearing slider bar and its internal structure follows the layered and sealed design of Harris and Bush \cite{Harris2015a}. Although precise alignment of the fluid bath can be achieved using the three aforementioned posts, we found it to be much easier to rapidly align the bath by tilting the whole optical table. We do this by first over inflating and then releasing air from the four pressurised optical table legs while remotely monitoring (using a tablet connected with the level via Bluetooth interface) the readings of the two-axis digital level. It is straightforward to reach the maximum accuracy of $\pm 0.002$ degrees of our measuring device this way.

To measure and characterize the bath vibrations, we follow Harris and Bush \cite{Harris2015a} and use two piezoelectric single-axis accelerometers (PCB 352C65) each positioned $60$ mm radial distance from the bath centre. The accelerometers have a sensitivity of $102$ mV/g, which we use for converting measured voltages to acceleration, and are supplied with a $4$ mA constant current excitation from a four-channel line-powered signal conditioner (PCB 482C05). The sensor voltages from the accelerometers are read in by a data acquisition device (NI USB-6343) connected to a computer, and also by a digital storage oscilloscope. 

\subsection{Droplet printer}\label{printer}
To introduce the droplets onto the surface of the liquid, we have constructed a droplet printer comprising a droplet generator mounted onto a two-axis linear translation stage. We have assembled a computer controlled, lead screw driven, stage (OpenBuilds C-Beam XY Actuator) to enable fast, precise, and repeatable deposition of many droplets in arbitrary planar configurations onto the fluid bath. We use a CNC Shield v3.0 mounted onto an Arduino Uno R3 (ATmega328P) microcontroller for both driving the two stepper motors (NEMA 23) that control the planar position of the droplet generator, and for generating the voltage pulses to expel the droplets. Sophisticated printing procedures could be programmed using the industry standard open source Grbl software. Since our intention is to eventually operate the entire experiment via Python scripts and the labscript suite \cite{Starkey2013a}, we have chosen to explicitly program the required voltage pulses onto the microcontroller to drive the stepper motors.

The droplets are produced using a piezoelectric droplet generator mounted at the end of one of the two printer arms. Our droplet generator is inspired by and builds upon the cost effective design of Harris and Bush \cite{Harris2015b,Ionkin2018a}. The fluid chamber is printed out of nylon (PA11) using a 3D printer (HP Jet Fusion) and it uses the same $35$~mm diameter piezoelectric buzzer disks as in Refs~\cite{Harris2015b,Ionkin2018a} clamped onto the fluid chamber. The \texttt{.stl} file of our 3D printed droplet printer is available in the Supplement \cite{supplement}. We use widely available replaceable M6 threaded 3D printer nozzles made of brass with a $0.1-1.0$~mm nozzle size in our droplet generator. We use the H-bridge circuitry of the third axis stepper motor driver available in the CNC Shield to send similar square voltage pulses as in \cite{Harris2015b} with a range of $\pm$ 30 V to the piezoelecric disk that expels the droplets. As explained in Refs.~\cite{Harris2015b,Ionkin2018a}, the nozzle-size-dependent pulse duration, typically on the order of a millisecond, must be tuned to achieve clean droplet production. We use a peristaltic pump and a micrometer translation stage to adjust the height of the fluid in the reservoir, which has a similar design as in Refs.~\cite{Harris2015b,Ionkin2018a}. The fluid level in the reservoir sets the hydrostatic pressure at the end of the nozzle to achieve clean and repeatable droplet ejection. The pump and the fluid reservoir are mounted rigidly onto the optical table, and the latter is connected to the moving droplet generator by a flexible silicone tubing (ID 2.0~mm, OD 4.0~mm). To avoid fluid pressure variations in the system due to the motion of the fluid filled tube, we ensure it does not impact any parts of the instrument during the motion of the printer arm.

According to the specifications sheet of the linear actuator its positioning accuracy is $91\;\mu$m, well below the typical size of the millimetric droplets. It is also straightforward to mount multiple droplet generators to simultaneously cater for a broad range of droplet sizes, or additional tools, such as directional launchers or optical instruments, onto the droplet printer arm. In light of the ease at which the droplets can be repeatably deposited in precise computer controlled locations on the fluid bath, leads us to anticipate that such droplet printers will be widely adopted becoming an indispensable component of any driven droplet experiment.  

\subsection{Imaging}\label{imaging}
The relatively slow (up to 50 mm/s) planar motion of the droplets can be tracked using a machine vision (top view imaging) camera (Allied Vision Manta G-158, Fujinon 1.5MP 12.5mm C Mount Lens) with up to $70$~fps frame rate. The image stream is acquired using a Python script with bindings for Open CV, which provides powerful particle tracking functionality.

The fast vertical bouncing motion (side view imaging) of the droplets is achieved using a high-speed camera (Krontech Chronos 1.4 38,565 fps maximum frame rate) and a microscope lens with 100 mm working distance. Since the droplet bouncing occurs in a typical experiment at a $40$~Hz frequency the high-speed camera enables a clean detection and determination of the droplet bouncing modes. The high-speed camera is rigidly mounted on a computer controlled linear translation stage, which allows the whole camera to be moved to enable precise focusing and focus stacking. The translation stage is mounted on a height and tilt adjustable support structure constructed from intersecting steel posts (Thorlabs P14) that are rigidly attached on the optical table.

In addition to the two fixed `quantitative' imaging systems, we occasionally use high resolution cameras readily available in mobile devices for `qualitative' imaging for the purposes of visualization and rapid prototyping.

\subsection{Control software and data acquisition}\label{control}
Presently, our experiment is controlled distributedly by several subsystems. Following Harris and Bush \cite{Harris2015a}, we use a PC that runs LabView to interface with the data acquisition device (NI-USB 6343) for generating the driving signal for the shaker and for reading in the accelerometer data. Both read and write operations are performed at $32\;$ kHz sampling rate. The analog signals from the two accelerometers are also monitored and analysed using a digital storage oscilloscope. A software feedback loop running once per second is sufficient to maintain a fixed driving amplitude, which would otherwise be drifting over long periods of time. 

Once the programmed driving signal of the bath is stabilised, the droplets are deposited onto the fluid surface using the droplet printer. A desired droplet deposition pattern is programmed onto the microcontroller that drives the droplet printer. A switch button is used for starting the printer and once the print job finishes a homing cycle returns the printer in its initial position where it waits idle until manually triggered to repeat. A second microcontroller is dedicated for monitoring the resistances of the two platinum temperature probes installed in the instrument.

In this work, all imaging was conducted by manually triggering the cameras, and the acquired image files were also post processed manually. Nevertheless, all of our subsystems have provisions to be integrated and be controlled by a single workstation allowing us to implement hardware timed Python scripted experiments using the labscript suite control software \cite{Starkey2013a}.

\begin{figure}
\centering
\includegraphics[width=\columnwidth]{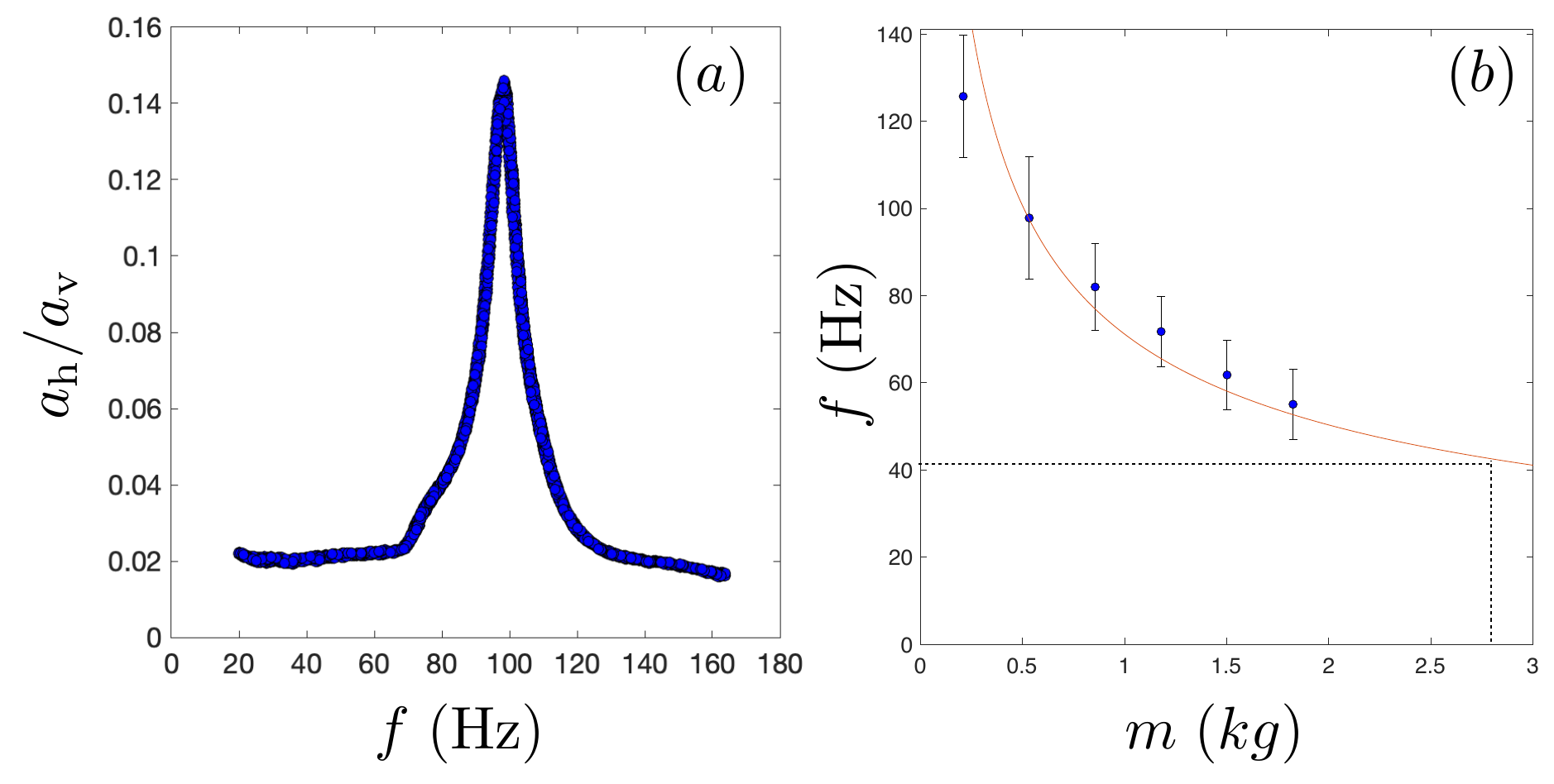}
\caption{
Mechanical unstabilised shaker resonances. (left) Ratio of horizontal to vertical acceleration as a function of the vertical driving frequency $f$ for a payload $m = 0.533$ kg. (right) Resonance frequency $\omega_0/(2\pi)$ as a function of the test mass $m$. The error bars are the widths $\pm \Gamma$ of the fitted Lorentzian line shapes. The vertical driving acceleration was held fixed at $a_{ 0}=400$~mV / (
102~mV/g). The dash-dotted lines show the predicted shaker resonance for the designed payload of the experiment.
}
\label{fig:shakertest}
\end{figure}

\section{Shaker performance}\label{shakertest}

Following the procedures of Harris and Bush \cite{Harris2015a}, we performed a baseline measurement on the plain shaker to characterise its mechanical resonance properties. The accelerometers were mounted onto a circular aluminium disk, which was attached to the shaker using machine bolts yielding a total minimum shaker payload of $m_{\rm mount} =0.210$ kg. To study payload dependence of the shaker resonances, we deployed five $5$~mm thick circular steel disks of radius $R_{\rm disk}=105$~mm, each weighing $m_{\rm disk} =0.323$ kg and stacked in between the shaker and the accelerometer mount. Figure \ref{fig:shakertest}(left) shows the ratio of the horizontal to vertical acceleration as a function of the sinusoidal driving frequency $f$ for a fixed payload. The line shape and amplitude of this resonance conforms with the expectation based on the results of Harris and Bush \cite{Harris2015a}. The solid curve is a fit to a Lorenzian line shape
\begin{equation}
g(\omega) = \frac{1}{2\pi}\frac{\Gamma^2/4}{(\omega-\omega_0)^2 +\Gamma^2/4},
\end{equation}
where $\omega=2\pi f$ is the angular frequency and the line width $\Gamma$ and resonance frequency $\omega_0$ are fit parameters.

Figure~\ref{fig:shakertest}(right) shows the resonance frequency $\omega_0/(2\pi)$ as a function of the payload mass $m$. The error bars correspond to the linewidths $\Gamma$. The solid curve shows a simple harmonic oscillator resonance $\omega_0=\sqrt{k/m}$, where $k = 0.2\times10^6$~N/m is the spring constant. These results are in good agreement with the observations of Harris and Bush \cite{Harris2015a} for the same shaker model.

\begin{figure}[!t]
\centering
\includegraphics[width=\columnwidth]{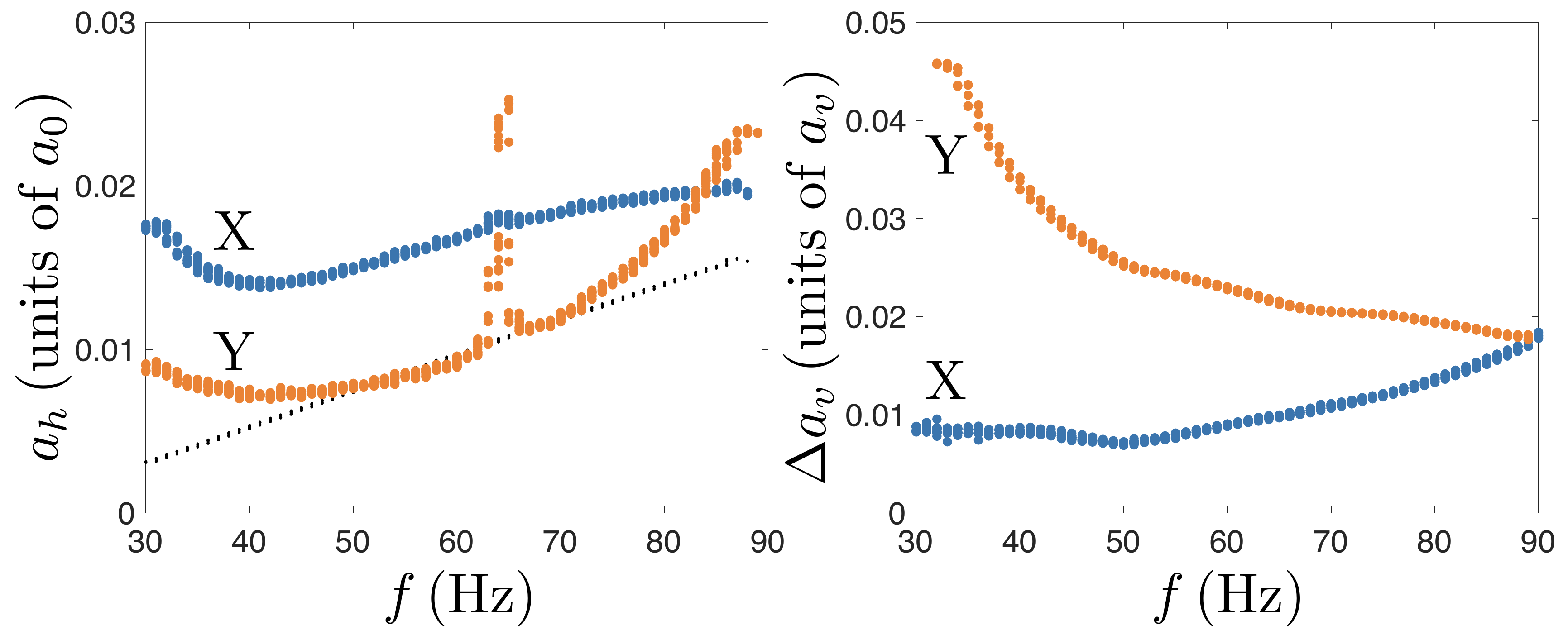}
\caption{
Mechanical airbearing stabilised shaker resonances. (left) Ratio $a_{h} /a_{v}$ of horizontal to vertical acceleration as a function of the vertical driving frequency $f$. The horizontal line is the reading when the shaker is turned off and the dotted line is scaled by a factor $0.02$ and shows the vertical driving acceleration of the bath. (right) Ratio $\Delta a_{v} /a_{v}$ of vertical acceleration difference between two accelerometers to the mean vertical acceleration as a function of the vertical driving frequency $f$. In both frames measurements are conducted along two Cartesian axes X and Y.  
}
\label{fig:airbearingtest}
\end{figure}

\subsection{Airbearing stabilization}\label{bearingtest}

As documented in the previous literature \cite{Harris2015a} and demonstrated above, the bare electrodynamic shaker suffers from mechanical resonances that compromise the purity of uniaxial oscillations. The transverse vibrations can be greatly reduced by stabilising the system with an air bearing that allows near frictionless motion in the axial direction while its stiff pressurised air spring prevents motion in the transverse plane. The air bearing is coupled to the shaker via a thin drive rod, which helps to further decouple transverse motion of the shaker from propagating through the air bearing \cite{Harris2015a}.

Nevertheless, care must be taken when designing the system since air bearings are vulnerable to pneumatic hammer instabilities, which limit the maximum useable payload and achievable stiffness (maximum supply pressure of air). As noted by Harris and Bush \cite{Harris2015a}, by choosing an air bearing with a large surface area and keeping the total payload (slider bar plus bath) to a minimum without compromising other design criteria such as the vibrational rigidity of the fluid bath, allows the shaker resonances to be effectively suppressed.

Figure \ref{fig:airbearingtest} (left) shows a baseline measurement with the full payload mass of $3150$~g, including the air bearing and the fluid bath but excluding fluid. The horizontal accelerations $a_h$ were measured along two Cartesian directions (X,Y). The Y trace reveals an air bearing pneumatic hammer resonance that for the considered payload occurs for multiples of $66$~Hz. The dashed line provides a reference for the used vertical driving amplitude of the shaker, which approximately follows the functional dependence of the Faraday threshold on the driving frequency.  The horizontal solid line shows the accelerometer reading when the shaker is turned off.

The plain shaker resonance that according to Fig.~\ref{fig:shakertest} would be expected to appear around $40$~Hz is well attenuated by the air bearing and the peak horizontal vibrations have been significantly reduced. However, the bottleneck has moved from the shaker to the air bearing which is now limiting the uniformity of the vibrations and the maximum useable payload.

Figure \ref{fig:airbearingtest}(right) shows vertical compliance of our system. The two traces correspond to measurements for which the accelerometers are in line along one of the two Cartesian axes. Although the vertical non-uniformity of the vibrations is only a few percent, it is still an order of magnitude larger than  achieved by Harris and Bush \cite{Harris2015a}. The reason for this is that our air bearing is susceptible to a pneumatic hammer instability, which remains the limiting factor in our current system. Nevertheless, our system's performance can be significantly improved by shortening the air bearing slider bar, which in our present system is $200$ mm longer than the air bearing carriage causing unnecessary torque that lowers the threshold for the pneumatic hammer instability. Due to the associated financial risk, we have chosen to postpone such a modification for the time being. Even better solution would be to acquire an air bearing with a larger surface area that has been shown to completely eliminate both the shaker resonance and pneumatic hammer resonance issues \cite{Harris2015a}.


\begin{figure}[t]
\centering
\includegraphics[width=0.95\columnwidth]{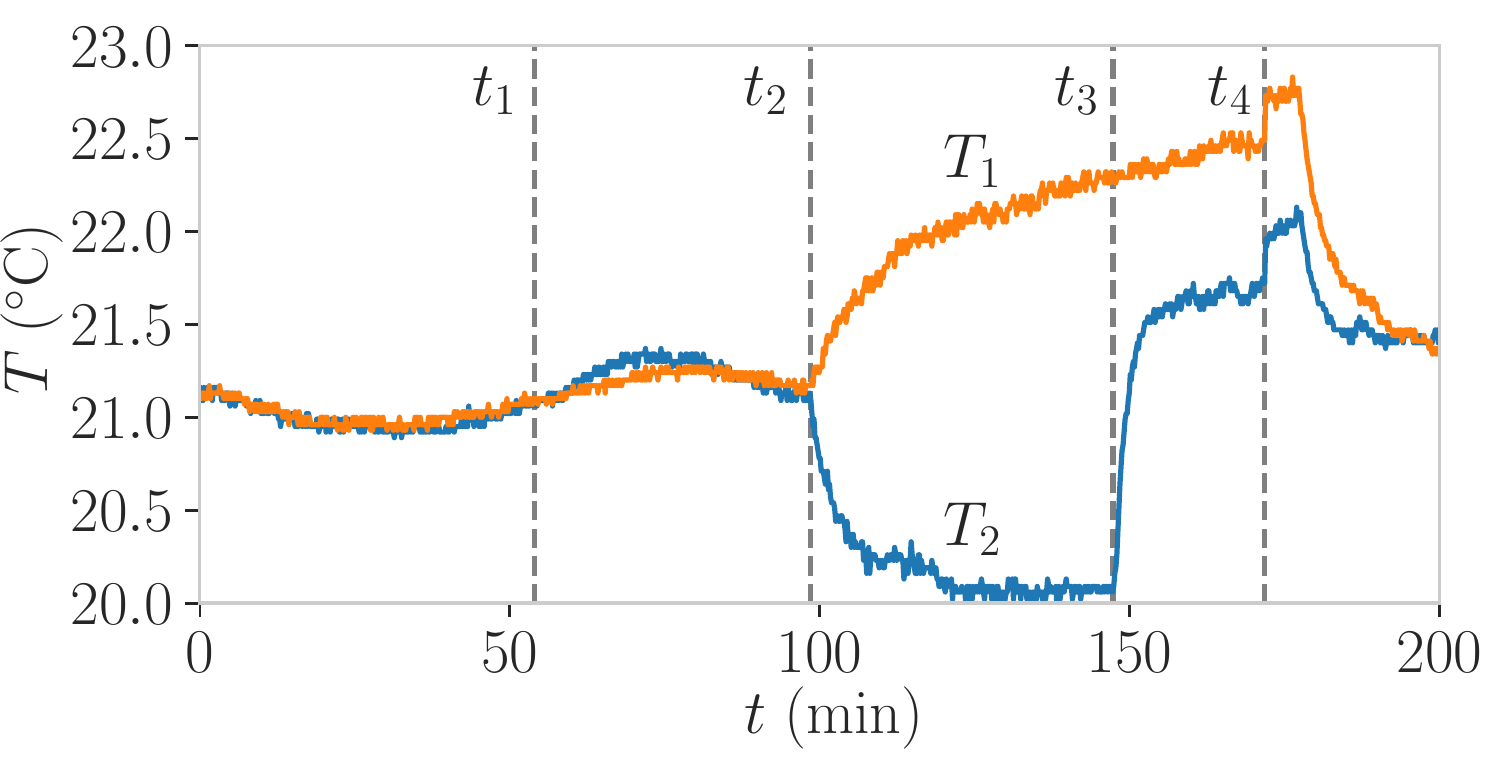}
\caption{
Temperature reading of thermometers $T_1$ (orange) and $T_2$ (blue) as functions of time. The conditions of the environment have been deliberately changed at times $t_i$ as described in the text.
}
\label{fig:temppi}
\end{figure}

\subsection{Thermal characterisation}\label{thermal}

\begin{figure}
\centering
\includegraphics[width=0.9\columnwidth]{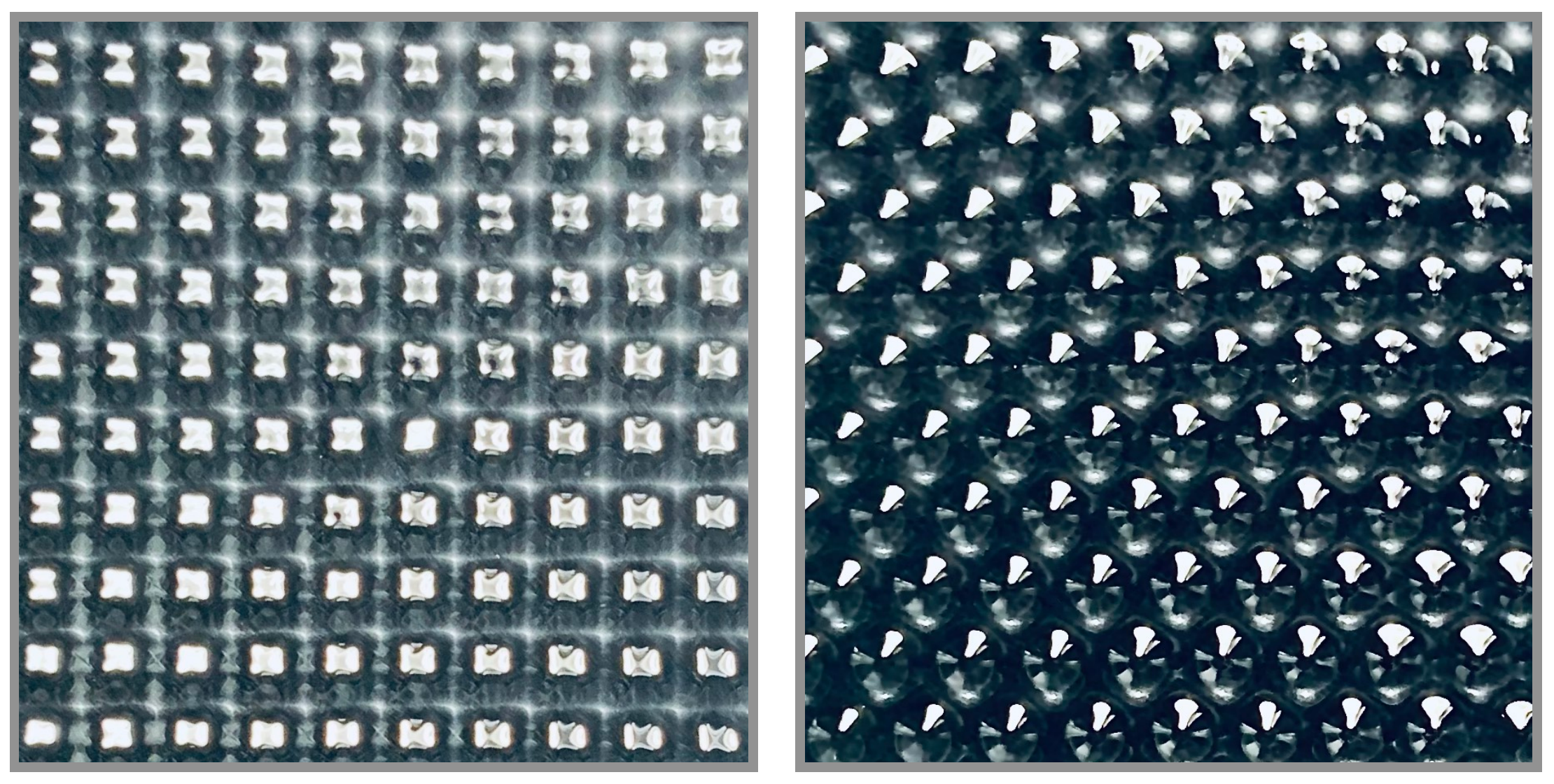}
\caption{
Faraday waves on the surface of a layer of silicone oil. Left image shows a snapshot of a square space and time crystal Faraday wave oscillating at a frequency of 40 Hz produced by driving the bath with a single 80 Hz sine wave with a 500 mV peak excitation, corresponding to $\Gamma_{80}=1.15 \Gamma^{\rm F}_{80}$. The nearest peak distance (lattice constant) 4.6~mm of this pattern is consistent with the expected 4.75~mm Faraday wave length \cite{MolaBush2013a,Tambasco2018a}.  
The triangular lattice (right) was produced using a two frequency ($f=80$~Hz and $f=40$~Hz) sine wave driving with a 600 mV peak excitation. The driving amplitude ratio of the waves $\Gamma_{40}/\Gamma_{80}=0.15$ and the two waves are phase shifted by $\pi/4$. This pattern has a nearest peak distance (lattice constant) of 5.4~mm. Movies S1 and S2 visualize the dynamics of these Faraday waves \cite{supplement}.
}
\label{fig:faraday}
\end{figure}

Our laboratory is fitted with an air conditioning system intended to be able to maintain a 0.5$^\circ$ C temperature stability within the laboratory environment. We use two PT100 platinum RTD probes with MAX31865 temperature sensor amplifiers and an ATmega328P microcontroller, to monitor the temperatures of the ambient air and the fluid. Instead of using the ice point as absolute reference, we calibrate the thermometers relative to each other and set the reference value to the 21$^\circ$ C ambient temperature of the laboratory. We use Sigma--Aldrich silicone oil having density $\rho=950$~kg/m$^3$ and viscosity of $20$ cSt at 25$^\circ$ C for the experiments. 

In principle, precision measurement, or active control \cite{Ellegaard2020a}, of the fluid temperature acting as a proxy of the fluid viscosity enables converting the measured absolute acceleration of the fluid bath to acceleration relative to the Faraday threshold that sets an absolute reference point intrinsic to the physical system. In practice, this turns out to be unnecessary for our  setup due to the good temperature stability in our laboratory, unless experiments requiring extreme precision are to be conducted. The thermal characterisation is summarised in Fig.~\ref{fig:temppi} that shows temperature of the fluid ($T_1$ orange) and the air ($T_2$ blue) as functions of time. The tip of the $T_1$ temperature probe is immersed into the fluid and the $T_2$ probe is placed 23~cm away from the centre of the bath, at the same vertical height. 

Until $t=t_1$ all equipment remain turned off and the temperature variation in Fig.~\ref{fig:temppi} reflects the ambient temperature fluctuations in the laboratory. At $t=t_1$ the shaker and the associated appliances are powered up yet the temperature remains within the specifications of the air conditioning system. This means that it is possible to operate the experiment continuously for long periods of time while maintaining steady thermal environment for the experiments.

At $t=t_2$ a 135 W (Aputure 120d) LED imaging light, placed 50 cm away from the bath, is turned on at maximum power providing illumination for high-speed imaging. The $T_1$ probe is directly exposed to the LED photons and detects a rising temperature. The $T_2$ probe is in a shadow behind an aluminium beam and shows dropping temperature, likely due to the response of the air conditioner that works to maintain the average temperature of the laboratory at its specified value. 

At $t=t_3$ the $T_2$ probe is brought to light from the dark side of the aluminium extrusion and consequently its reading jumps up. In the interval $[t_3-t_4]$ the difference in the readings of $T_1$ and $T_2$ is due to the position of $T_2$ being offset away from the line of sight from the LED beam and thereby being exposed to lower light intensity. 
\begin{figure*}[!ht]
\centering
\includegraphics[width=1.9\columnwidth]{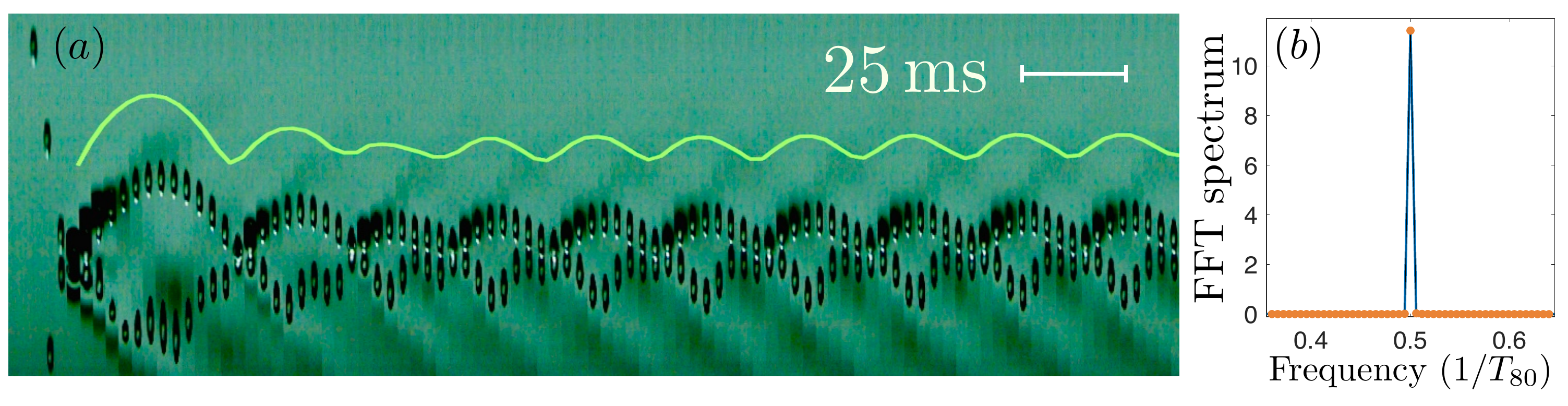}
\caption{
Liquid time crystal behaviour. Visualization of the first ten bounces of a droplet dropped $12$~mm above the fluid surface, is shown in (a) together with the height function $h(t)$ (solid curve) extracted by tracking the brightest pixel in each video frame. The Fourier power spectrum $|\mathcal{F}(h)|^2$ with $T_{80}=1/f$ of the next one hundred bounces is shown in (b). The droplet bounced for more than one hundred thousand times before it was annihilated by the operator of the experiment. The early time evolution of the droplet is shown in the Movie S3 \cite{supplement}.
}
\label{fig:tc}
\end{figure*}

At $t=t_4$ the temperature first jumps up slightly due to the experimenter walking around the optical table to generate turbulent air currents in the laboratory. Shortly after this the LED lighting is turned off at the time of the peak temperature reading, and the measured temperatures rapidly drop back to the ambient laboratory temperature.

We conclude that continuously measuring the fluid temperature in day to day experiments does not provide a practical means to improve on the results of typical droplet experiments unless they are particularly sensitive to the Faraday threshold. However, the spot heating due to the intense illumination from the imaging light generates turbulent convection air currents, in addition to those due to cooling fans, the air bearing, and moving experimentalists, and have the ability to significantly affect the dynamics of the droplets. It is for this reason that the lighting required by high-speed imaging should only be turned on when necessary. Nevertheless, the adverse effects of turbulent air currents can also be eliminated by protecting the fluid bath with enclosures instead of keeping the bath open.

\section{Creation and observation of liquid time crystals}\label{results}

Having characterised the design features of the instrument, we deploy it to create and observe liquid time crystal behaviours of gravitationally bouncing wave-droplet entities. The key characteristic of discrete time crystals is the spontaneous breaking of discrete time translation symmetry and the emergence of time-periodic dynamics in the system at a frequency lower than the driving \cite{Sacha2018a,Else2019a}. The time crystal concept was introduced to experiments exhibiting period multiplying phenomena in Refs.~\cite{Zhang2017a,Choi2017a,Rovny2018a}. Since then, time crystals have featured in numerous experiments, including Bose--Einstein condensates and Fermi superfluids \cite{Smits2018a,Autti2021a,Seman2021a,Hemmerich2021a}. Here we shed the time crystal perspective onto the periodically driven droplet systems.

\subsection{Faraday waves}\label{faraday}

When planar layer of fluid is vibrated vertically at a sufficiently high, frequency dependent, amplitude the fluid surface becomes unstable and develops patterns of Faraday waves \cite{Faraday1931a,Kumar1994a}. Expressing the vertical acceleration of the fluid bath as 
\begin{equation}
    a_v(f,t) = \Gamma_f g\sin(2\pi f t),
    \label{driving}
\end{equation}
where $\Gamma_f$ is the frequency dependent dimensionless amplitude and $g$ is the acceleration due to Earth's gravity, the Faraday waves emerge at and above the Faraday threshold $\Gamma_f=\Gamma_f^{\rm F}$. The value of the Faraday threshold is strongly dependent on the driving frequency $f$ and the fluid viscosity. While the driving frequency can be set and controlled precisely, the fluid viscosity is a more challenging control parameter due to it being dependent on the temperature in addition to the intrinsic properties of the fluid.

We have determined the Faraday threshold for a plain $f=80$~Hz sinusoidal driving of the form Eq.~(\ref{driving}) by visually inspecting the fluid surface as the amplitude $\Gamma_{80}$ is varied at the stable 21$^\circ$ C ambient temperature, and at the elevated temperature just after $t_4$. In both cases when the Faraday threshold is rapidly crossed in both directions the Faraday waves are seen to emerge for $440\pm 10$ mV peak excitation, corresponding to $\Gamma^{\rm F}_{80}= 4.3\pm 0.1$. This compares well with the theory prediction of Mol{\'a}{\v{c}}ek and Bush \cite{MolaBush2013a} who calculated Faraday threshold $\Gamma^{\rm F}_{80}=4.115$ for the $20$~cSt silicone oil in the infinitely deep bath approximation, as well as the experimental result of $\Gamma^{\rm F}_{80}= 4.25$ by Couchman and Bush \cite{couchman_bush_2020}. To be able to detect the shift in the Faraday threshold due to the ambient temperature variations would require precision measurement of the fluid surface height variations, in addition to precision control and stabilization of the external conditions. 

When the system is driven significantly above the Faraday threshold, the Faraday waves rapidly grow in amplitude and are then easy to observe. The observable Faraday patterns that include squares, triangles, and quasicrystals, have been thoroughly discussed in the previous literature, such as in Refs~\cite{Faraday1931a,Kumar1994a,Edwards1994a}. Figure \ref{fig:faraday} shows Faraday patterns with a square (left) and triangular (right) unit cell, produced using our instrument. These patterns repeat at a frequency of $40$~Hz, whereas the primary driving frequency of the fluid bath is $80$~Hz. Movies S1 and S2, respectively, show a side view perspective to the Faraday wave dynamics of these patterns.

Faraday waves have also been created in Bose--Einstein condensates \cite{Engels2007a} and more recently discussed in the context of discrete time crystals \cite{Smits2018a,Seman2021a}. The life time, quantified by the number of time crystal periods, in such superfluid systems has been on the order of one hundred. By contrast, in the case of driven room temperature silicone oil the Faraday wave time crystal pattern in Fig.~(\ref{fig:faraday}) repeats with a period of $2/f$ for all practical purposes indefinitely.

\subsection{Gravitationally bouncing droplets}\label{bouncing}

Already when the bath acceleration remains far below the Faraday threshold, droplets that begin to bounce may be introduced onto the fluid surface \cite{Couder2005a,Couder2006a,Walker1978a}. Simple spring models have been shown to capture well the essential features of the resulting bouncing dynamics \cite{MolaBush2013a,WindWillasen2013a,Bush2020a}.

For given intrinsic fluid properties and external ambient conditions three important tuneable parameters to specify are the droplet size $R_d$, driving frequency $f$, and amplitude constant $\Gamma_f$ of the bath acceleration. There exists a rather large volume of this three dimensional parameter space where droplets may stably bounce in the so called $(2,1)$ bouncing mode, where we do not make a distinction between the long and short contact bouncers \cite{MolaBush2013a,WindWillasen2013a,Tambasco2018a}. In this particular bouncing mode, the centre of mass of the droplet undergoes vertical periodic oscillations at the frequency $f/2$ touching the fluid surface once every two oscillation periods of the fluid bath that is driven at frequency $f$. We note that in addition to the vertical centre-of-mass motion of the droplets, they also support internal vibrational modes in free space with a peculiar dispersion relation \cite{Rayleigh1879a,Steen2019a}
\begin{equation}
\omega_n =  \sqrt{\sigma/\rho} \;k_n^{3/2},
\label{dropdisp}
\end{equation}
which is also predicted to hold for quantum droplets \cite{HuiLiu2020a} comprised of self-trapped Bose--Einstein condensates \cite{Pfau2021a}. In Eq.~(\ref{dropdisp}) $k_n= [n(n-1)(n+2)]^{1/3}/R_d$, $\sigma$ is the surface tension, and $n$ is an integer.

We first fix the bath oscillation frequency to $f=80$ Hz and the peak excitation to $270$ mV and then drop a droplet onto the bath using the droplet generator. The nozzle diameter was $0.5$~mm and the duration of the voltage pulse was 1.5~ms. The droplet settles into a stable bouncing state in a matter of few bounces, as shown in Fig.~(\ref{fig:tc})(a) and thereafter keeps bouncing coherently, see Fig.~(\ref{fig:tc})(b), for exceedingly long periods of time. We have followed a droplet bouncing at $40$~Hz for over $N=10^5$ bounces (45 minutes) at which point we manually terminated the droplet. We wish to draw attention to the extreme stability and life time, measured in the number of bouncing periods, of these liquid time crystal droplets. By carefully protecting them from the specs of laboratory dust and turbulent air currents should enable them to keep bouncing for many hours if not days. 

Whereas the Faraday wave time crystals extend throughout the space, the bouncing droplet time crystals are spatially localized to within the volume of space occupied by the droplet. We also note that it is straightforward to create bouncing droplets on demand in many different bouncing modes where the droplet bouncing period is longer than the driving period of the bath \cite{MolaBush2013a,WindWillasen2013a,Bush2020a}. Furthermore, it is straightforward to initialize the droplet in a phase where it is resonantly bouncing at the driving frequency $f$. By smoothly increasing the driving amplitude while keeping all other parameters fixed, may induce period multiplying bifurcations allowing continuous observations of the emergence of the discrete time crystal phases such as the one represented by Fig.~\ref{fig:tc}. These gravitationally bouncing droplets are closely related to the time crystals predicted to emerge if a cloud of Bose--Einstein condensed atoms is made to gravitationally bounce off an oscillating sheet of laser light \cite{Sacha2015a,Giergiel2020a}. The bouncing droplet time crystals establish points of comparison for experiments investigating quantum and classical effects in time crystals.

\begin{figure}
\centering
\includegraphics[width=0.9\columnwidth]{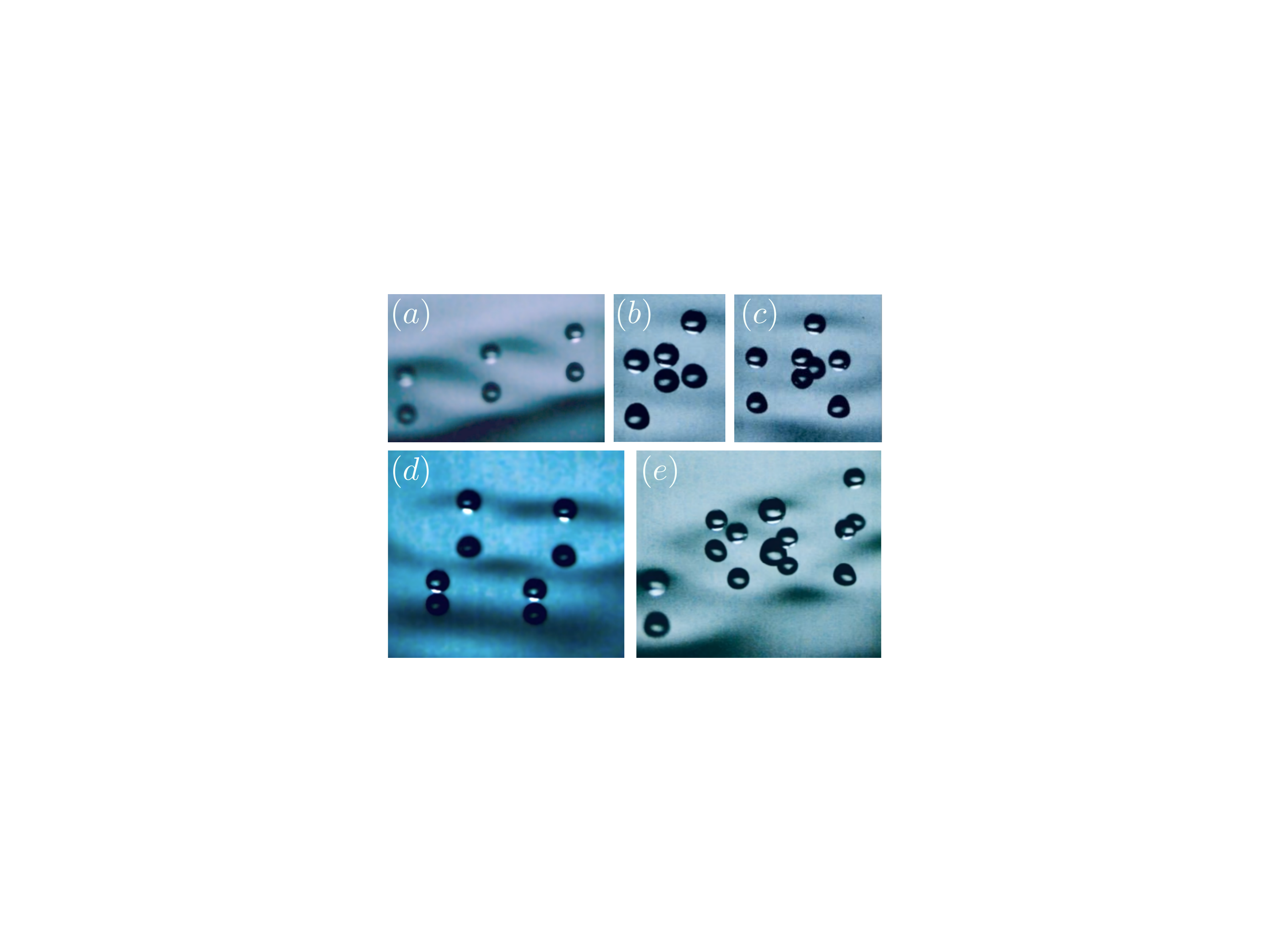}
\caption{
Interacting time and space crystals showing (a) an out-of-focus image of a one-space-dimensional ferromagnetic time crystal, (b) one-space-dimensional antiferromagnetic time crystal, (c) two-space-dimensional triangular antiferromagnetic time crystal, (d) two-space-dimensional square antiferromagnetic time crystal, and (e) disordered impurity doped time crystal. Note that in each image twice the number of actual droplets are visible due to the mirror reflection from the fluid surface. For the dynamics of these time crystals see the Movies S4-S7 \cite{supplement}.
}
\label{fig:tcexamples}
\end{figure}

By depositing multiple droplets onto the fluid we may create a large variety of interacting time crystals. Figure~\ref{fig:tcexamples}(a) shows a linear array of in-phase bouncing droplets that constitute a one-space-dimensional `ferromagnetic' time crystal. Figure~\ref{fig:tcexamples}(b) shows a time crystal with antiferromagnetic order, which can be converted to a two-space-dimensional antiferromagnetic triangular time crystal Fig.~\ref{fig:tcexamples}(c) by adding one more droplet to the system. We may also create square time crystals as shown in Fig.~\ref{fig:tcexamples}(d). Figure~\ref{fig:tcexamples}(e) shows spatially disordered, impurity doped, time crystal with an imbalanced droplet size containing in total seven droplets with five small ones, one medium size, and one large droplet. The structures in Figs~\ref{fig:tcexamples}(a)-(e) are stabilised by (time crystal)-(time crystal) interactions. The dynamics of these structures is clarified by the appended movies S3-S7 \cite{supplement}.

In addition to the spatial crystal structures, it is straightforward to generate arbitrary temporal driving signals \cite{Perrard2016a,Valani2019a} that allow realization of designer time crystal potentials. This makes it possible to access a broad class of classic condensed matter phenomena in the time-domain, ranging from disorder driven localization effects to topological states of time crystal matter \cite{Sacha2018a}.

\subsection{Many-body dynamics}\label{manybody}

The droplet printer-generator allows creating arbitrary two-dimensional patterns of droplets in the bath. This is particularly useful for experiments, such as the spin-lattice systems \cite{Saenz2018a,Saenz2021a} that use fixed sub-surface topographies for trapping the droplets in specific positions that may be selectively loaded using the droplet printer. However, it is also possible to produce self-organized structures of many interacting droplets in the absence of external confining `potentials' where the resulting structures are determined by the droplet-droplet interactions.
 
\begin{figure}
\centering
\includegraphics[width=\columnwidth]{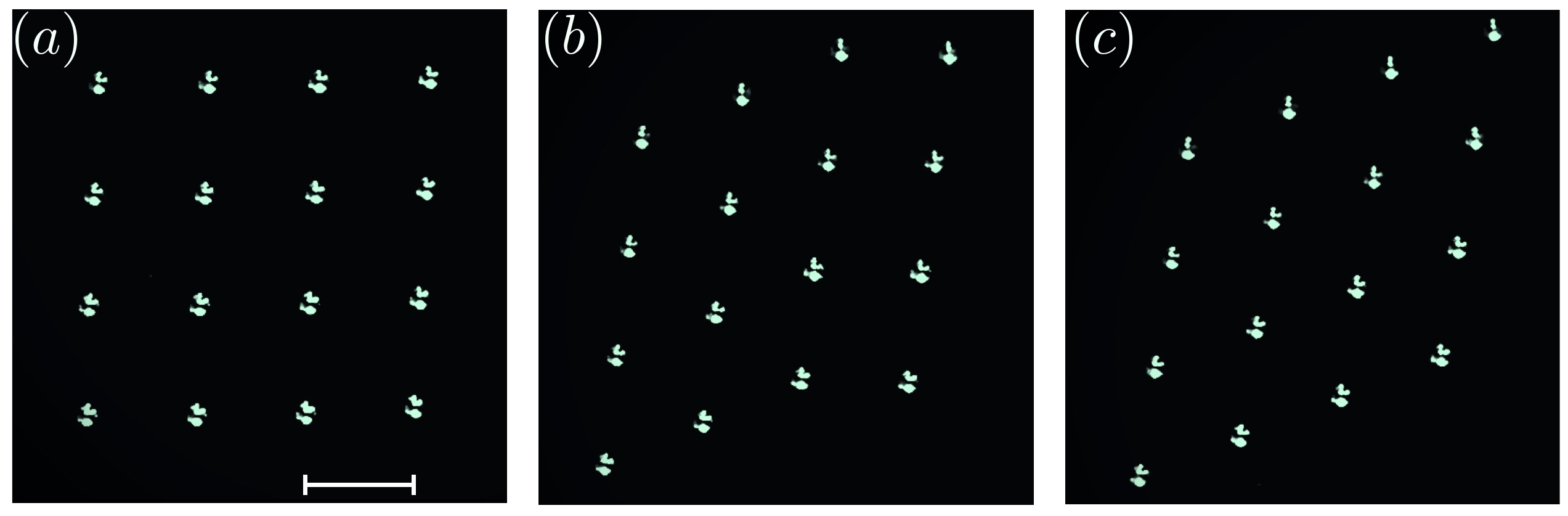}
\caption{
Spontaneous transmutation of a square lattice to a triangular lattice. 
Image (a) shows sixteen droplets printed into a long-lived metastable square lattice pattern. Image (b) shows an intermediate state where a square and triangle arrangements co-exist. Image (c) shows a stable triangle lattice after the transmutation is complete. Images (a), (b), and (c) were acquired 187, 293, and 328 seconds after the square lattice was printed. The scale bar in (a) has length $7.7\;{\rm mm}$ and applies to all frames, and the droplets were generated using a 1.0~mm nozzle diameter and 2.3~ms pulse width. The full dynamics of this transformation is shown in the movie S8 that runs at twice its acquisition speed \cite{supplement}.
}
\label{fig:square2tria}
\end{figure}

Figure~\ref{fig:square2tria}(a) shows a square lattice of sixteen superwalker droplets \cite{Valani2019a,Valani2021a} three minutes after having been deposited onto the bath using the droplet printer generator. The bath is driven with a superposition of two sine waves of the form Eq.~(\ref{driving}) with frequencies $f_1=80$~Hz and $f_2=40$~Hz and an amplitude ratio $\Gamma_{40}/\Gamma_{80}=0.05$ and a $\pi/8$ phase offset. The peak excitation is set to 270 mV, which is significantly lower than the 440 mV that would be required to excite Faraday waves.

Movie S8 \cite{supplement} shows the production and the subsequent evolution of the droplets. After approximately five minutes the square lattice undergoes a transmutation to a triangle lattice as shown in Fig.~\ref{fig:square2tria} (b) and (c). The triangle lattice remains a stable configuration thereafter. This sequence of events is robust and repeatable, although the time scales vary significantly from run to run due to turbulent air currents that we allow to excite crystal oscillations akin to phonons \cite{Eddi2009a,Eddi2011a}. In comparison to the experiments by Thomson, Couchman and Bush who observed free droplet rings deforming to polygonal shapes by increasing the amplitude of the driving \cite{couchman_bush_2020,Thomson2020a}, here the deformation to triangular lattice occurs spontaneously from the long-lived metastable square lattice. An instability of a square lattice to a triangular lattice was also mentioned by Eddi and coworkers for droplets with single frequency driving \cite{Eddi2009a,Eddi2011a}.

In condensed matter systems, competition between square and triangular lattice configurations are ubiquitous. For instance, Abrikosov found a square vortex flux lattice solution for Type II superconductors \cite{Abrikosov1957a} and it turns out that a triangle lattice is preferred with only a 2\% lower free energy \cite{Kleiner1964a}. However, it is not obvious that an equilibrium concept such as free energy could be meaningfully applied to driven non-equilibrium systems that are not subject to energy conservation in the first place. As such, the square to triangle transmutation illustrated in Fig.~\ref{fig:square2tria} is a rather non-trivial phenomenon, even when viewed from the perspective of the Floquet frame. 

While it might be tempting to explain these observations in terms of droplet-free Faraday patterns such as those in Fig.~\ref{fig:faraday}, such interpretation is challenged by the fact that the system is driven far from the Faraday threshold and that upon increasing the driving amplitude we first observe circular, then square and only after that triangular Faraday wave patterns.

Yet another attempt to explain the preference toward a triangle lattice is that it yields higher circle packing fraction than a square lattice. However, this alone cannot explain the observations without a deeper understanding of why the droplets should prefer arrangements with higher packing fractions. Thus we conclude by attributing the observed behaviour to the complex and self-consistent interplay due to wave-mediated many-body interactions between the droplets and subtle boundary effects \cite{Edwards1994a}, and that a detailed analysis, beyond the scope of this study, will be warranted to provide a quantitative understanding for the observed phenomenon.

\section{discussion}\label{conclusions}

We have described our new instrument for conducting experiments on non-equilibrium physics of driven droplets. An air bearing was used for stabilizing the electrodynamic shaker resonances. Care must be taken in choosing the air bearing or else the shaker resonance is removed only to be replaced by a pneumatic hammer instability of the air bearing as the limiting factor for achieving uniform and uniaxial vibrations.

We have described a versatile integrated droplet printer-generator, which is the key novelty of our instrument. The printer allows for rapid on-demand computer controlled production of arbitrary two-dimensional droplet patterns enabling a systematic exploration of free space droplet lattices, and fast loading of spin-lattices \cite{Saenz2018a,Saenz2021a}. The ability to automate the droplet production opens many new opportunities and makes it feasible to efficiently study larger ensembles to bring down statistical uncertainties. 

We deployed our instrument to print long-lived liquid time crystals comprised of the periodically driven droplets, gravitationally bouncing at half the driving frequency. We have followed such time crystals for over one hundred thousand oscillation periods. Our current setup does not enable us to verify the phase coherence of these time crystals for over such long time scales as it would require an ultra precise time reference. However, it would be possible for us to import an atomic clock timing signal from our neighbouring laboratory to provide us such a capability. As a technology demonstration, we also presented a gallery of one-dimensional, two-dimensional, ferromagnetic, antiferromagnetic, and impurity doped disordered many-body liquid time crystals, and observed long-lived metastable square lattices and their transmutations to arrays with triangular symmetry. We hope our results will provide useful points of comparison, thereby helping to distinguish between classical and quantum behaviours in time crystal experiments.

\begin{acknowledgements}
I am grateful to John Bush, Jeff Davis, Peter Hannaford, Kris Helmerson, Sascha Hoinka, Krzysztof Sacha, Rahil Valani and Chris Vale for useful discussions, Paul Cahill, Dan Kapsaskis and Jonathan Tollerud for laboratory management, Aidan O'Keeffe for mechanical workshop services, and Mark Whitehead for laser cutting and 3D printing services. The computer aided designs were produced using Shapr3D. This research was funded by the Australian Government through the Australian Research Council (ARC) Future Fellowship FT180100020.
\end{acknowledgements}

\bibliographystyle{apsrev4-1}

\end{document}